\documentclass[onecolumn]{article}%
\usepackage{amsmath}
\usepackage{amssymb}
\usepackage{amsfonts}
\usepackage{graphicx}%
\providecommand{\U}[1]{\protect\rule{.1in}{.1in}}
\newtheorem{theorem}{Theorem}[section]

\newtheorem{corollary}[theorem]{Corollary}

\newtheorem{example}[theorem]{Example}

\newtheorem{lemma}[theorem]{Lemma}

\newtheorem{proposition}[theorem]{Proposition}

\newenvironment{proof}[1][Proof]{\noindent\textbf{#1.}}{\ \rule{0.5em}{0.5em}}
\newenvironment{theorem-nonum}[1][Theorem]{\noindent\textbf{#1.} }{\ \rule{0.5em}{0.5em}}
\begin{document}

\title{Reverse Test and Characterization of Quantum Relative Entropy}
\author{Keiji Matsumoto
\and National Institute of Informatics, 2-1-2, Hitotsubashi, Chiyoda-ku, Tokyo\\Quantum Computation and Information Project, SORST, JST,\\5-28-3, Hongo, Bunkyo-ku, Tokyo 113-0033, Japan}
\maketitle

\begin{abstract}
The aim of the present paper is to give axiomatic characterization of quantum
relative entropy utilizing resource conversion scenario. We consider two sets
of axioms: non-asymptotic and asymptotic. In the former setting, we prove that
the upperbound and the lowerbund of $\mathrm{D}^{Q}\left(  \rho||\sigma
\right)  $ is $\mathrm{D}^{R}\left(  \rho||\sigma\right)  :=\mathrm{tr}%
\,\rho\ln\sqrt{\rho}\sigma^{-1}\sqrt{\rho}$ and $\mathrm{D}\left(
\rho||\sigma\right)  :=$ $\mathrm{tr}\,\rho\left(  \ln\rho-\ln\sigma\right)
$, respectively. In the latter setting, we prove uniqueness of quantum
relative entropy, that is, $\mathrm{D}^{Q}\left(  \rho||\sigma\right)  $
should equal a constant multiple of $\mathrm{D}\left(  \rho||\sigma\right)  $.
In the analysis, we define and use reverse test and asymptotic reverse test,
which are natural inverse of hypothesis test.

\end{abstract}

\section{Introduction}

Many problems in quantum/classical information theory can be viewed as
conversion between given resources and 'standard' resources, and such
viewpoint had turned out to be very fruitful. This manuscript will
exploit\ this scenario in asymptotic theory of quantum estimation theory (with
some comments on classical estimation theory). Resource conversion scenario
was first explored in axiomatic theory of entanglement measures. The optimal
asymptotic conversion ratio from maximally entangled states (`standard'
resource) to a given state is called \textit{entanglement cost}, while the
optimal ratio for inverse conversion is called \textit{distillable
entanglement}. It had been shown that all quantities which satisfies a set of
reasonable axioms takes value between these two quantities. Similar argument
had been applied to classical/quantum channels, and so on.

The aim of the present paper is to give axiomatic characterization of quantum
relative entropy utilizing resource conversion scenario. We consider two sets
of axioms: non-asymptotic and asymptotic. In both cases, we require a quantum
relative entropy $\mathrm{D}^{Q}\left(  \rho||\sigma\right)  $ is monotone
decreasing by application of any CPTP map. In addition, in the former setting,
we assume quantum relative entropy coincide with its classical counterpart for
probability distributions $\left\{  p,q\right\}  $: $\mathrm{D}^{Q}\left(
p||q\right)  =\mathrm{D}\left(  p||q\right)  $. Then we can prove that the
upperbound and the lowerbound of $\mathrm{D}^{Q}\left(  \rho||\sigma\right)  $
is
\[
\mathrm{D}^{R}\left(  \rho||\sigma\right)  :=\mathrm{tr}\,\rho\ln\sqrt{\rho
}\sigma^{-1}\sqrt{\rho},
\]
and
\[
\mathrm{D}\left(  \rho||\sigma\right)  :=\mathrm{tr}\,\rho\left(  \ln\rho
-\ln\sigma\right)  ,
\]
respectively. In the latter setting, in stead, $\mathrm{D}^{Q}\left(
\rho||\sigma\right)  $ is supposed to satisfy some asymptotic properties,
namely weak additivity and lower asymptotic continuity, which will be defined
later. Under such assumptions, we prove uniqueness of quantum relative
entropy, that is, $\mathrm{D}^{Q}\left(  \rho||\sigma\right)  $ should equal a
constant multiple of $\mathrm{D}\left(  \rho||\sigma\right)  $.

In the analysis, newly defined \textit{\ reverse test} and
\textit{\ asymptotic reverse test} play key role. The former is a conversion
from a pair $\left\{  p,q\right\}  $\textrm{\ }of probability distributions to
a pair $\left\{  \rho,\sigma\right\}  $ of quantum states, and the latter is
an approximate conversion from a pair $\left\{  p^{n},q^{n}\right\}  $ of
probability distributions over the binary set $\left\{  0,1\right\}  $ to a
pair $\left\{  \rho^{n},\sigma^{n}\right\}  $ of quantum states. Each of them
is natural inverse of optimal measurement for hypothesis and hypothesis test
of Neyman-Pearson type, and optimal measurement for hypothesis test, respectively.

In the course of analyzing reverse test, we show operational meaning of RLD
Fisher information. Also, we prove joint convexity of $\mathrm{D}^{R}\left(
\rho||\sigma\right)  $.

\section{Main results}

In the paper, the totality of density operators in the Hilbert space
$\mathcal{H}$ is denoted by $\mathcal{S}\left(  \mathcal{H}\right)  $, and the
totality rank $r$ elements is denoted by $\mathcal{S}_{r}\left(
\mathcal{H}\right)  $. Unless otherwise mentioned, we suppose $d:=\dim
\mathcal{H}<\infty$. \ We consider following conditions.

\begin{description}
\item[(M)] (Monotonicity) For any CPTP map $\Lambda$,
\[
\mathrm{D}^{Q}\left(  \rho||\sigma\right)  \geq\mathrm{D}^{Q}\left(
\Lambda\left(  \rho\right)  ||\Lambda\left(  \sigma\right)  \right)  .
\]

\item[(N)] (Normalization) For any probability distributions $\left\{
p,q\right\}  $,
\[
\mathrm{D}^{Q}\left(  p||q\right)  =\mathrm{D}\left(  p||q\right)  :=\sum
_{x}p\left(  x\right)  \left(  \ln p\left(  x\right)  -\ln q\left(  x\right)
\right)  .
\]

\item[(A)] (Weak additivity)
\[
\mathrm{D}^{Q}\left(  \rho^{\otimes n}||\sigma^{\otimes n}\right)
=n\mathrm{D}^{Q}\left(  \rho||\sigma\right)
\]

\item[(C)] (Lower asymptotic continuity) \
\begin{equation}
\lim_{n\rightarrow\infty}\left\Vert \tilde{\rho}^{n}-\rho^{\otimes
n}\right\Vert =0\Longrightarrow\varliminf_{n\rightarrow\infty}\frac{1}%
{n}\left\{  \mathrm{D}^{Q}\left(  \tilde{\rho}^{n}||\sigma^{\otimes n}\right)
-\mathrm{D}^{Q}\left(  \rho^{\otimes n}||\sigma^{\otimes n}\right)  \right\}
\geq0. \label{asym-lower-cont}%
\end{equation}

\end{description}

Define
\begin{align*}
\mathrm{D}\left(  \rho||\sigma\right)   &  :=\mathrm{tr}\,\rho\left(  \ln
\rho-\ln\sigma\right)  ,\\
\mathrm{D}^{R}\left(  \rho||\sigma\right)   &  :=\mathrm{tr}\,\rho\ln
\sqrt{\rho}\sigma^{-1}\sqrt{\rho},
\end{align*}
and denote by $M\left(  \rho\right)  $ the probability distribution of the
data from the application of the measurement $M$ to $\rho$.

\begin{theorem}
\label{th:DM<DQ<DR}If (M) and (N) are satisfied,
\[
\max_{M}\mathrm{D}\left(  M\left(  \rho\right)  ||M\left(  \sigma\right)
\right)  \leq\mathrm{D}^{Q}\left(  \rho||\sigma\right)  \leq\mathrm{D}%
^{R}\left(  \rho||\sigma\right)  .
\]

\end{theorem}

\begin{theorem}
\label{th:D<DQ<DR}If (M), (N) and (A) are satisfied,
\[
\mathrm{D}\left(  \rho||\sigma\right)  \leq\mathrm{D}^{Q}\left(  \rho
||\sigma\right)  \leq\mathrm{D}^{R}\left(  \rho||\sigma\right)  .
\]

\end{theorem}

\begin{theorem}
\label{th:DQ=D}If (M), (A), and (C) are satisfied,
\[
\mathrm{D}^{Q}\left(  \rho||\sigma\right)  =const.\times\mathrm{D}\left(
\rho||\sigma\right)  .
\]

\end{theorem}

\subsection{Proof of Main theorems}

Below, \textit{reverse test} \ of a pair of states $\left\{  \rho
,\sigma\right\}  $ means the triplet $\left(  \Phi,\left\{  p,q\right\}
\right)  $ of a CPTP map $\Phi$ and probability distributions $p$, $q$ with
\[
\Phi\left(  p\right)  =\rho,\,\Phi\left(  q\right)  =\sigma.
\]
We use following theorems to prove main theorems:

\begin{theorem}
\label{th:reverse-test} \
\[
\min\mathrm{D}\left(  p||q\right)  =\mathrm{D}^{R}\left(  \rho||\sigma\right)
,
\]
where minimization is taken for over all reverse tests $\left(  \Phi,\left\{
p,q\right\}  \right)  $ of $\left\{  \rho,\sigma\right\}  $.
\end{theorem}

\begin{theorem}
\label{th:Hiai-Petz}(Hiai-Petz\thinspace\cite{HiaiPetz}) For any states $\rho$
and $\sigma$, and constant $c>0$, we can find a projective measurement
$M^{n}:=\left\{  P^{n},\mathbf{1}-P^{n}\right\}  $ such that:
\begin{align*}
\lim_{n\rightarrow\infty}\mathrm{tr}\,P^{n}\rho^{\otimes n}  &  =1,\,\\
\lim\frac{-1}{n}\ln\mathrm{tr}\,P^{n}\sigma^{\otimes n}  &  \geq
\mathrm{D}(\rho||\sigma)-c,\\
\lim_{n\rightarrow\infty}\frac{1}{n}\mathrm{D}(M^{n}\left(  \rho^{\otimes
n}\right)  \,||\,M^{n}\left(  \sigma^{\otimes n}\right)  )  &  \geq
\mathrm{D}(\rho||\sigma)-c.
\end{align*}

\end{theorem}

\begin{proposition}
\label{prop:D-asym-cont}$\mathrm{D}\left(  \rho||\sigma\right)  $ satisfies
the condition (C).
\end{proposition}

\begin{proof}
By Fannes's inequality, when $n$ is very large,%
\begin{align*}
&  \frac{1}{n}\left\vert \mathrm{D}\left(  \rho^{\otimes n}||\sigma^{\otimes
n}\right)  -\mathrm{D}\left(  \tilde{\rho}^{n}||\sigma^{\otimes n}\right)
\right\vert \\
&  \leq\left\Vert \rho^{\otimes n}-\tilde{\rho}^{n}\right\Vert _{1}\ln
d+\frac{1}{n}\\
&  +\left\Vert \rho^{\otimes n}-\tilde{\rho}^{n}\right\Vert _{1}\left\{
\text{the first eigenvalue of }\ln\sigma\right\} \\
&  \rightarrow0.
\end{align*}

\end{proof}

\begin{theorem}
\label{th:asym-q-template} If $\mathrm{D}\left(  \rho_{0}||\sigma_{0}\right)
>\mathrm{D}\left(  \rho||\sigma\right)  $, there is a sequence $\left\{
\Psi^{n}\right\}  $ of TPCP map with
\begin{equation}
\lim_{n\rightarrow\infty}\left\Vert \Psi^{n}\left(  \rho_{0}^{\otimes
n}\right)  -\rho^{\otimes n}\right\Vert _{1}=0,\,\,\,\Psi^{n}\left(
\sigma_{0}^{\otimes n}\right)  =\sigma^{\otimes n}.\label{asym-def}%
\end{equation}
Conversely, if such $\left\{  \Psi^{n}\right\}  $ with (\ref{asym-def})
exists, $\mathrm{D}\left(  \rho_{0}||\sigma_{0}\right)  \geq\mathrm{D}\left(
\rho||\sigma\right)  $.
\end{theorem}

Proof of Theorem\thinspace\ref{th:reverse-test} and Theorem\thinspace
\ref{th:asym-q-template} will be given later in Subsection\thinspace
\ref{subsec:reverse-test} and Subsection\thinspace
\ref{subsec:asym-cont-monotone-relative-entropy}\ , respectively. Here, we use
these to prove the main theorems.

\textbf{Proof \ of Theorem\thinspace}\ref{th:DM<DQ<DR}\textbf{. \ \ }\ \ That
$\mathrm{D}^{R}$ satisfies (M) is known \cite{HiaiPetz}, but here we give
another proof. By Theorem\thinspace\ref{th:reverse-test},
\begin{align*}
&  \mathrm{D}^{R}(\Lambda\left(  \rho\right)  ||\Lambda\left(  \sigma\right)
)=\min\left\{  \,\mathrm{D}\left(  p||q\right)  \,;\,\Phi\left(  p\right)
=\Lambda\left(  \rho\right)  ,\,\Phi\left(  q\right)  =\Lambda\left(
\sigma\right)  \right\} \\
&  \leq\min\left\{  \mathrm{D}\left(  p||q\right)  \,;\left.  \Phi=\Psi
\circ\Lambda,\text{ s.t., }\Psi\left(  p\right)  =\rho,\,\Psi\left(  q\right)
=\sigma\right.  \right\} \\
&  =\min\left\{  \mathrm{D}\left(  p||q\right)  \,;\text{ }\Psi\left(
p\right)  =\rho,\,\Psi\left(  q\right)  =\sigma\right\} \\
&  =\mathrm{D}^{R}(\rho||\sigma).
\end{align*}
So $\mathrm{D}^{R}$ satisfies (M). (N) is obviously satisfied. Also, that
$\max_{M}\mathrm{D}\left(  M\left(  \rho\right)  ||M\left(  \sigma\right)
\right)  $ satisfies (M) and (N) is trivial.

Letting $(\Phi,\{p,q\})$ be an optimal reverse test, due to (N) and (M),
\begin{align*}
\mathrm{D}^{R}(\rho||\sigma)  &  =\mathrm{D}(p||q)\underset{\text{(N)}}%
{=}\mathrm{D}^{Q}(p||q)\\
\underset{\text{(M)}}{\geq}\mathrm{D}^{Q}(\Phi(p)||\Phi(q))  &  =\mathrm{D}%
^{Q}(\rho||\sigma).
\end{align*}
Also,%
\[
\mathrm{D}(M\left(  \rho\right)  ||M\left(  \sigma\right)  )\underset
{\text{(N)}}{=}\mathrm{D}^{Q}(M\left(  \rho\right)  ||M\left(  \sigma\right)
)\underset{\text{(M)}}{\leq}\mathrm{D}^{Q}(\rho||\sigma).
\]
${\tiny \blacksquare}$

\textbf{Proof \ of Theorem\thinspace}\ref{th:D<DQ<DR}\textbf{. \ }Since
$\mathrm{D}^{R}(\rho||\sigma)$ is weakly additive, we have the upper bound.
The lower bound is known\thinspace\cite{Hayashi:2005}, but also can be easily
obtained by Theorem\thinspace\ref{th:DM<DQ<DR} and Theorem\thinspace
\ref{th:Hiai-Petz} . ${\tiny \blacksquare}$

\textbf{\ }

\textbf{Proof \ of Theorem\thinspace}\ref{th:DQ=D}\textbf{. \ }Without loss of
generality, we can suppose%
\[
\mathrm{D}^{Q}\left(  \rho_{0}||\sigma_{0}\right)  =\mathrm{D}\left(  \rho
_{0}||\sigma_{0}\right)  ,
\]
for some $\rho_{0}$, $\sigma_{0}$. \ For given $\rho$ and $\sigma$, let $l$,
$l^{\prime}$, $m$, $m^{\prime}$ be integers with
\[
\frac{l^{\prime}}{m^{\prime}}\mathrm{D}\left(  \rho_{0}||\sigma_{0}\right)
<\mathrm{D}\left(  \rho||\sigma\right)  <\frac{l}{m}\mathrm{D}\left(  \rho
_{0}||\sigma_{0}\right)  .
\]
By Theorem\thinspace\ref{th:asym-q-template}, there is $\left\{  \Psi
^{n}\right\}  $ with%
\begin{align*}
\lim_{n\rightarrow\infty}\left\Vert \Psi^{n}\left(  \rho_{0}^{\otimes
l\,n}\right)  -\rho^{\otimes mn}\right\Vert _{1}  &  =0,\,\\
\Psi^{n}\left(  \sigma_{0}^{\otimes l\,n}\right)   &  =\sigma^{\otimes mn}.
\end{align*}
Since $\mathrm{D}^{Q}$ is satisfies (A), (C) , and (M), we have
\begin{align*}
&  m\mathrm{D}^{Q}\left(  \rho||\sigma\right)  \underset{\text{(A)}}%
{=}\varlimsup_{n\rightarrow\infty}\frac{1}{n}\mathrm{D}^{Q}\left(
\rho^{\otimes mn}||\sigma^{\otimes mn}\right) \\
&  \underset{\text{(C)}}{\leq}\varliminf_{n\rightarrow\infty}\frac{1}%
{n}\mathrm{D}^{Q}\left(  \Psi^{n}\left(  \rho_{0}^{\otimes l\,n}\right)
||\Psi^{n}\left(  \sigma_{0}^{\otimes l\,n}\right)  \right) \\
&  \underset{\text{(M)}}{\leq}\varliminf_{n\rightarrow\infty}\frac{1}%
{n}\mathrm{D}^{Q}\left(  \rho_{0}^{\otimes nl}||\sigma_{0}^{\otimes
nl}\right)  \underset{\text{(A)}}{=}l\mathrm{D}^{Q}\left(  \rho_{0}%
||\sigma_{0}\right)  ,
\end{align*}
or
\[
\mathrm{D}^{Q}\left(  \rho||\sigma\right)  \leq\frac{l}{m}\mathrm{D}%
^{Q}\left(  \rho_{0}||\sigma_{0}\right)  =\frac{l}{m}\mathrm{D}\left(
\rho_{0}||\sigma_{0}\right)  .
\]
Exchanging $\left\{  \rho_{0},\sigma_{0}\right\}  $ and $\left\{  \rho
,\sigma\right\}  $ in the above argument, we obtain
\[
\mathrm{D}^{Q}\left(  \rho||\sigma\right)  \geq\frac{l^{\prime}}{m^{\prime}%
}\mathrm{D}^{Q}\left(  \rho_{0}||\sigma_{0}\right)  =\frac{l^{\prime}%
}{m^{\prime}}\mathrm{D}\left(  \rho_{0}||\sigma_{0}\right)  .
\]
Taking $\frac{l}{m}$ and $\frac{l^{\prime}}{m^{\prime}}$ arbitrarily close to
$\frac{\mathrm{D}\left(  \rho||\sigma\right)  }{\mathrm{D}\left(  \rho
_{0}||\sigma_{0}\right)  }$, we have
\[
\mathrm{D}^{Q}\left(  \rho||\sigma\right)  =\mathrm{D}\left(  \rho
||\sigma\right)  .
\]
${\tiny \blacksquare}$\textbf{\ \ }

\section{Monotone metric}

\subsection{Classical Fisher Information as a monotone metric}

Let us consider a family of probability distribution $\left\{  \text{
}p_{\theta};\theta\in\Theta\subset%
\mathbb{R}
^{m}\right\}  $ over the finite set $\mathcal{X}$, $\left\vert \mathcal{X}%
\right\vert <\infty$. A logarithmic derivative is defined by $l_{\theta
,i}:=\partial_{i}\ln p_{\theta}$, where $\partial_{i}:=\frac{\partial
}{\partial\theta^{i}}$. Fisher information $J_{\theta}$ (sometimes denoted as
$J_{p_{\theta}}$) is defined by
\[
J_{\theta,i,j}:=\sum_{x}p_{\theta}(x)l_{\theta,i}\left(  x\right)
l_{\theta,j}\left(  x\right)  =\sum_{x}l_{\theta,i}\left(  x\right)
\partial_{j}p_{\theta}(x).
\]
It is known that, with some regularity condition, the optimal asymptotic mean
square error of an estimate of $\theta$ equals $J_{\theta}^{-1}$.

Being positive definite and covariant by the coordinate change of the
parameter space, $J_{\theta}$ induces a Riemannian metric, or an inner product
in the tangent space $\mathcal{T}_{\theta}$ by
\[
J_{\theta}\left(  \partial_{i}p_{\theta},\partial_{j}p_{\theta}\right)
:=J_{\theta,i,j}\,,
\]
where the representation of $\mathcal{T}_{\theta}$ is chosen as $\mathrm{span}%
\left\{  \partial_{i}p_{\theta};i=1,\cdots,m\right\}  $. This metric brings
about the following intuitive picture: the precision of estimate is
proportional to the distance between $p_{\theta}$ and $p_{\theta
+\mathrm{d}\theta}$. .

Hereafter,the differential map of affine map $\Lambda$ is also denoted by
$\Lambda$, by abusing the notation. Cencov \cite{Cencov} had proven :

\begin{theorem}
\cite{Cencov} Suppose a Riemannian metric$\,\,g_{p_{\theta}}$ is monotone
decreasing by application of Markov maps,%
\[
g_{p_{\theta}}\left(  X,X\right)  \geq g_{\Lambda\left(  p_{\theta}\right)
}\left(  \Lambda\left(  X\right)  ,\Lambda\left(  X\right)  \right)  .
\]
Then, $g_{p_{\theta}}$ is the one induced by Fisher information, up to a
constant multiple.
\end{theorem}

In the proof, it is essential that the metric is Riemannian, i.e., the norm in
the tangent space is defined via an inner product. This assumption can be
replaced by weak additivity and asymptotic lower continuity
\cite{Matsumoto:2010}.

\subsection{SLD and RLD Fisher information}

We consider a family $\left\{  \rho_{\theta};\theta\in\Theta\subset%
\mathbb{R}
^{m}\right\}  $ of density operators, and suppose the map $\theta
\rightarrow\rho_{\theta}$ $\ $is smooth enough, and $\Theta$ is open. Define a
\textit{symmetric logarithmic derivative} (SLD) $L_{\theta,i}^{S}$and a
\textit{right logarithmic derivative} (RLD) $L_{\theta,i}^{R}$ as a solution
to the matrix equation,%
\[
\partial_{i}\rho_{\theta}=\frac{1}{2}(L_{\theta,i}^{S}\rho_{\theta}%
+\rho_{\theta}L_{\theta,i}^{S})=L_{\theta,i}^{R}\rho_{\theta}.
\]
If $\rho_{\theta}$ is strictly positive, $L_{\theta,i}^{S}$ and $L_{\theta
,i}^{R}$ are uniquely defined in this way. If $\rho_{\theta}$ has zero
eigenvalues, $L_{\theta,i}^{S}$ still can be defined, but not uniquely.
$L_{\theta,i}^{R}$ exists (and if exists, unique) if and only if $\partial
_{i}\rho_{\theta}$ has non-zero eigenvalues only in the support of
$\rho_{\theta}$. Observe they are quantum equivalences of a classical
logarithmic derivative, $l_{\theta,i}=\partial_{i}\ln p_{\theta}(x)$.

SLD Fisher information matrix $J_{\theta}^{S}$ and RLD Fisher information
matrix $J_{\theta}^{R}$ is defined as%
\[
J_{\theta,i,j}^{S}=\Re\mathrm{Tr}\rho_{\theta}L_{\theta,i}^{S}L_{\theta,j}%
^{S},\,\,J_{\theta,i,j}^{R}=\mathrm{Tr}\rho_{\theta}L_{\theta,j}^{R\dagger
}L_{\theta,i}^{R},
\]
respectively \cite{Holevo}. They are quantum analogues of classical Fisher
information $J_{\theta}$, and, being positive definite, each of \ them induces
inner product to the tangent space $\mathcal{T}_{\theta}$,
\[
J_{\theta}^{S}\left(  \partial_{i}\rho_{\theta},\partial_{j}\rho_{\theta
}\right)  :=J_{\theta,i,j}^{S}\,\,,\quad J_{\theta}^{R}\left(  \partial
_{i}\rho_{\theta},\partial_{j}\rho_{\theta}\right)  :=J_{\theta,i,j}^{R}\,,
\]
where we represent $\mathcal{T}_{\theta}$ by $\mathrm{span}\left\{
\partial_{i}\rho_{\theta};i=1,\cdots,m\right\}  $. We sometimes use notations
such as $J_{\rho_{\theta}}^{S}$ and $J_{\rho_{\theta}}^{R}$to indicate that
the underlying family of states is $\left\{  \rho_{\theta}\right\}  $. Even if
$\rho_{\theta}$ is not full-rank, $J_{\theta}^{S}$ is uniquely defined,
regardless the indefiniteness of SLD.

An operational meaning of SLD Fisher metric is given through estimation of
$\theta$ in an asymptotic setting, just like its classical counterpart. For
the detail, see, for example, \cite{Hayashi:book}. Here, we point out relation
of SLD Fisher information to classical Fisher information of the family
$\left\{  M\left(  \rho_{\theta}\right)  \right\}  $.

\begin{theorem}
\cite{Helstrom}\cite{Nagaoka}\cite{Hayashi:book}
\begin{equation}
J_{M\left(  \rho_{\theta}\right)  }\leq J_{\rho_{\theta}}^{S},\,
\label{sld-cr}%
\end{equation}
Also, for any $X\in$ $\mathcal{T}_{\theta}$, there is a measurement $M$ with%
\[
J_{M\left(  \rho_{\theta}\right)  }\left(  M\left(  X\right)  ,M\left(
X\right)  \right)  =J_{\rho_{\theta}}^{S}\left(  X,X\right)  .
\]

\end{theorem}

\subsection{RLD and Reverse SLD}

Denote by $\mathcal{W}$ the totality of matrices $W$ with $\mathrm{tr}%
\,WW^{\dagger}=1$. The totality of $d\times d^{\prime}$ elements of
$\mathcal{W}$ is denoted by $\mathcal{W}_{d^{\prime}}$, where $d^{\prime}\geq
r=\mathrm{rank\,}\rho$. \ Consider a map form $\mathcal{W}$ to $\mathcal{S}%
\left(  \mathcal{H}\right)  $ such that%
\[
W\rightarrow WW^{\dagger}.
\]
A meaning of this map is as follows. Let
\[
W=[\sqrt{p_{1}}\left\vert \phi_{1}\right\rangle ,\cdots,\sqrt{p_{d^{\prime}}%
}\left\vert \phi_{d^{\prime}}\right\rangle ],
\]
then,
\[
WW^{\dagger}=\sum_{i=1}^{d^{\prime}}p_{i}\left\vert \phi_{i}\right\rangle
\left\langle \phi_{i}\right\vert .
\]

\begin{proposition}
\label{th:A=BL}There is a Hermitian matrix $L$ with $A=BL$ if and only if
$AB^{\dagger}=BA^{\dagger}$ and $\operatorname{Im}A\subset\operatorname{Im}B$.
\end{proposition}

\begin{proof}
Since `\thinspace only if ' is trivial, we show ` if\thinspace'.Consider the
singular decomposition of $B$:
\[
B=UXV,
\]
where $U$ and $V$ is $d\times r$- and $r\times d^{\prime}$- matrix, with
$U^{\dagger}U=VV^{\dagger}=\mathbf{1}$, respectively. Let%
\[
L:=V^{\dagger}X^{-1}U^{\dagger}AB^{\dagger}UX^{-1}V+\tilde{V}^{\dagger}%
\tilde{V}C^{\dagger}V^{\dagger}V+V^{\dagger}VC\tilde{V}^{\dagger}\tilde{V}.
\]
Here $\tilde{V}$ is $\left(  d-r\right)  \times d^{\prime}$-matrix with
$\tilde{V}\tilde{V}^{\dagger}=\mathbf{1}$ and $\tilde{V}V^{\dagger}=0$, and
$C$ is a matrix with $A=BC$ (existence of such $C$ is due to
$\operatorname{Im}A\subset\operatorname{Im}B$) Since $AB^{\dagger}$ is
Hermitian, $L$ is Hermitian. Also,
\begin{align*}
BL  &  =UXV\left\{  V^{\dagger}X^{-1}U^{\dagger}A\left(  UXV\right)
^{\dagger}UX^{-1}V+\tilde{V}^{\dagger}\tilde{V}C^{\dagger}V^{\dagger
}V+V^{\dagger}VC\tilde{V}^{\dagger}\tilde{V}\right\} \\
&  =AV^{\dagger}V+UXV\tilde{V}^{\dagger}\tilde{V}C^{\dagger}V^{\dagger
}V+UXVC\tilde{V}^{\dagger}\tilde{V}\\
&  =AV^{\dagger}V+BC\tilde{V}^{\dagger}\tilde{V}\\
&  =A\left(  V^{\dagger}V+\tilde{V}^{\dagger}\tilde{V}\right)  =A\text{.}%
\end{align*}
Hence, $L$ satisfies required condition, and the assertion is proved.
\end{proof}

Since
\[
\,\left(  L_{\theta,i}^{R}W\right)  W^{\dagger}=L_{\theta,i}^{R}\rho_{\theta
}=\rho_{\theta}L_{\theta,i}^{R\dagger}=W\left(  L_{\theta,i}^{R}W\right)
^{\dagger},
\]
due to Proposition\thinspace\ref{th:A=BL},
\[
L_{\theta,i}^{R}W=WA_{\theta,i}^{R},\,\exists A_{\theta,i}^{R}=\left(
A_{\theta,i}^{R}\right)  ^{\dagger}.
\]
\bigskip\ $A_{\theta,i}^{R}$ is called the \textit{reverse SLD at }$W$.

On the other hand, let $A$ be an arbitrary $d^{\prime}\times d^{\prime}$
Hermitian matrix. Observe that the\ image of $WAW^{\dagger}$ is a subspace of
the image of $WW^{\dagger}$. Therefore, for an arbitrary reverse SLD, there is
a RLD, i.e.,%
\[
\forall A=A^{\dagger},\,\exists L^{R}\,\,\,L^{R}WW^{\dagger}=WAW^{\dagger},
\]
and, letting $Q$ be the projection onto $\left(  \ker W\right)  ^{\bot
}=\operatorname{Im}W^{\dagger}$, \
\[
\,\,\,L^{R}W=WAQ.
\]
(Especially, if $d^{\prime}=r$,$\,\,\,L^{R}W=WA.$) Therefore, we have, $.$%
\begin{align}
\mathrm{Tr}\rho L^{R\dagger}L^{R}  &  =\mathrm{Tr}L^{R}WW^{\dagger}%
L^{R\dagger}=\mathrm{Tr}WAQAW^{\dagger}\nonumber\\
&  \leq\mathrm{Tr}W\left(  A\right)  ^{2}W^{\dagger}. \label{AR>JR}%
\end{align}
Especially, if $d^{\prime}=r$, the equality holds.

\subsection{Reverse estimation of quantum state family and RLD}

The heart of quantum statistics is optimization of a measurement, i.e., choice
of a measurement which converts a family of quantum states to the most
informative classical probability distribution family. In estimation of the
parameter $\theta$ in asymptotic situation, we maximize the output Fisher
information $J_{M\left(  \rho_{\theta}\right)  }$ by modifying $M$.

Now, we consider the reverse of above, i.e., generation of the quantum state
family $\{\rho_{\theta}\}$: a pair $(\Phi,\{p_{\theta}\})$ is said to be a
reverse estimation of $\{\rho_{\theta}\}$ if
\[
\Phi\left(  p_{\theta}\right)  =\rho_{\theta},\,\,\forall\theta\in\Theta.
\]
Classical version of this is nothing but randomization criteria of deficiency,
the concept which plays key role in statistical decision theory\thinspace
\cite{Torgersen}. Let us introduce `local' version of this condition. We say
$\left(  \Phi,\left\{  p_{\theta},\partial_{i}p_{\theta};i=1,\cdots,m\right\}
\right)  $ is tangent reverse estimation of \bigskip$\left\{  \rho_{\theta
},\partial_{i}\rho_{\theta};i=1,\cdots,m\right\}  $ if
\begin{equation}
\Phi\left(  p_{\theta}\right)  =\rho_{\theta},\,\Phi\left(  \partial
_{i}p_{\theta}\right)  =\partial_{i}\rho_{\theta_{0}} \label{localsim}%
\end{equation}
hold at $\theta$. (In statistical decision theory, when this relation holds,
we say $\left\{  \rho_{\theta},\partial_{i}\rho_{\theta};i=1,\cdots,m\right\}
$ is locally deficient relative to$\left\{  p_{\theta},\partial_{i}p_{\theta
};i=1,\cdots,m\right\}  $ at $\theta\,$\cite{Torgersen}.)

Now let us consider the $m=1$-case, and optimize $\left(  \Phi,\left\{
p_{\theta},\mathrm{d}p_{\theta}/\mathrm{d}\theta\right\}  \right)  $ to
minimize the Fisher information $J_{p_{\theta}}$. Let us denote by $\delta
_{x}$ the delta-distribution at $x$. Suppose $\Phi\left(  \delta_{x}\right)  $
is pure (this can be supposed without loss of generality) and let
\[
\left\vert \phi_{x}\right\rangle \left\langle \phi_{x}\right\vert
:=\Phi\left(  \delta_{x}\right)  .
\]
Then (\ref{localsim}) is rewritten as%
\begin{align*}
\rho_{\theta}  &  =\sum_{x=1}^{d^{\prime}}p_{_{\theta}}(x)\left\vert \phi
_{x}\right\rangle \left\langle \phi_{x}\right\vert ,\\
\frac{\mathrm{d}\rho_{\theta}}{\mathrm{d}\theta}  &  =\sum_{x=1}^{d^{\prime}%
}\frac{\mathrm{d}p_{\theta}\left(  x\right)  }{\mathrm{d}\theta}\left\vert
\phi_{x}\right\rangle \left\langle \phi_{x}\right\vert .
\end{align*}
If $p_{_{\theta}}(x)=0$ and $\mathrm{d}p_{\theta}/\mathrm{d}\theta\neq0$, the
input Fisher information is infinite. So let us suppose this is not the case,
and let us define
\begin{align}
W  &  =[\sqrt{p_{\theta}(1)}\left\vert \phi_{1}\right\rangle ,\cdots
,\sqrt{p_{_{\theta}}(l)}\left\vert \phi_{d^{\prime}}\right\rangle ]\nonumber\\
A  &  =\mathrm{diag}\,\left(  \frac{1}{p_{\theta}(1)}\frac{\mathrm{d}%
p_{\theta}\left(  1\right)  }{\mathrm{d}\theta},\cdots,\frac{1}{p_{\theta
}(d^{\prime})}\frac{\mathrm{d}p_{\theta}\left(  d^{\prime}\right)
}{\mathrm{d}\theta}\right)  . \label{decom}%
\end{align}
Then, the input Fisher information is
\begin{equation}
\sum_{x=1}^{d^{\prime}}p_{_{\theta}}(x)\left(  \frac{1}{p_{\theta}(x)}%
\frac{\mathrm{d}p_{\theta}\left(  x\right)  }{\mathrm{d}\theta}\right)
^{2}=\mathrm{tr}\,WA^{2}W^{\dagger}\geq J_{\theta_{0}}^{R}, \label{J>JR}%
\end{equation}
where the inequality is due to (\ref{AR>JR}). \ 

On the other hand, let us suppose $\mathrm{rank}\,W=r$ and $W$ satisfies
$WW^{\dagger}=\rho_{\theta}$, and let $A$ be the reverse SLD at $W$,
$WAW^{\dagger}=\mathrm{d}\rho_{\theta}/\mathrm{d}\theta$. Then this $A$
achieves the equality of (\ref{AR>JR}). Since $\left(  WU\right)  \left(
WU\right)  ^{\dagger}=\rho_{\theta}$ for unitary matrix $U$, we can suppose
that $A$ is diagonal,by choosing $W$ properly. Therefore, one can define
$\Phi$ by tracking above process inverse way, which achieves identity of
(\ref{J>JR}). Therefore, we have:

\begin{theorem}
\label{th:r-rld-bound}Suppose $\dim\Theta=1$. Then
\begin{equation}
J_{p_{\theta}}\geq J_{\rho_{\theta}}^{R} \label{J>JR-2}%
\end{equation}
holds for all the reverse estimation of $\left\{  \rho_{\theta}\right\}  $,
and there is a tangent reverse estimation $\left(  \Phi,\left\{  p_{\theta
},\mathrm{d}p_{\theta}/\mathrm{d}\theta\right\}  \right)  $ with%
\[
J_{p_{\theta}}=J_{\rho_{\theta}}^{R}.
\]

\end{theorem}

$m>1$-case is briefly discussed in Appendix\thinspace
\ref{appendix:multi-reverse-est}.

\subsection{Monotone metric}

In this subsection, to avoid notational complexity, we let $m=1$, and
abbreviate $J_{\rho_{\theta}}^{S}\left(  \mathrm{d}\rho_{\theta}%
/\mathrm{d}\theta,\mathrm{d}\rho_{\theta}/\mathrm{d}\theta\right)  $ as
$J_{\rho_{\theta}}^{S}$, and so on. Corresponding statement for $m>1$-case
will be easily obtained by considering its appropriate one dimensional subfamily.

It is known that SLD Fisher metric and RLD Fisher metric are monotone
decreasing by application of CPTP\ maps
\[
J_{\rho_{\theta}}^{R}\geq J_{\Lambda\left(  \rho_{\theta}\right)  }%
^{R},\,\quad J_{\rho_{\theta}}^{S}\geq J_{\Lambda\left(  \rho_{\theta}\right)
}^{S},
\]
and any monotone Riemanian metric $g$, if a constant factor is properly
chosen, takes values between SLD and RLD Fisher metric
\[
J_{\rho_{\theta}}^{S}\leq g_{\rho_{\theta}}\leq J_{\rho_{\theta}}^{R},
\]
\cite{Petz}. In this section, we show the operational proof of the slightly
stronger version of these facts.

First, monotonicity of SLD is trivial because the optimization of measurement
applied to the family $\{\Lambda\left(  \rho_{\theta}\right)  \} $ is
equivalent to the optimization of measurement applied to $\{\rho_{\theta}\}$
over the restricted class of measurements of the form $M\circ\Lambda$:
\[
J_{\Lambda\left(  \rho_{\theta}\right)  }^{S}=\max_{M}J_{M\circ\Lambda\left(
\rho_{\theta}\right)  }\leq\max_{M}J_{M\left(  \rho_{\theta}\right)  }%
=J_{\rho_{\theta}}^{S}%
\]
The monotonicity of RLD Fisher metric is proven in the similar manner. Given a
tangent reverse estimation $(\Phi,\{p_{\theta},\mathrm{d}p_{\theta}%
/\mathrm{d}\theta\})$ of $\{\rho_{\theta},\mathrm{d}\rho_{\theta}%
/\mathrm{d}\theta\}$, $(\Lambda\circ\Phi,\{p_{\theta}\})$ is a tangent reverse
estimation of $\{\Lambda\left(  \rho_{\theta}\right)  ,\Lambda\left(
\mathrm{d}\rho_{\theta}/\mathrm{d}\theta\right)  \}$. Since $\{\Lambda\left(
\rho_{\theta}\right)  ,\Lambda\left(  \mathrm{d}\rho_{\theta}/\mathrm{d}%
\theta\right)  \}$ may have a better tangent reverse estimation, we have
\begin{align*}
J_{\Lambda\left(  \rho_{\theta}\right)  }^{R}  &  =\min\left\{  J_{p_{\theta}%
}\,;\,\Phi\left(  p_{\theta}\right)  =\Lambda\left(  \rho_{\theta}\right)
,\Phi\left(  \mathrm{d}p_{\theta}/\mathrm{d}\theta\right)  =\Lambda\left(
\mathrm{d}\rho_{\theta}/\mathrm{d}\theta\right)  \right\} \\
&  \leq\min\left\{  J_{p_{\theta}};\,\Phi=\,\Psi\circ\Lambda\,\text{,
\thinspace s.t. }\Psi\left(  p_{\theta}\right)  =\rho_{\theta},\,\Psi\left(
\mathrm{d}p_{\theta}/\mathrm{d}\theta\right)  =\mathrm{d}\rho_{\theta
}/\mathrm{d}\theta\right\} \\
&  =J_{\rho_{\theta}}^{R}.
\end{align*}

Assume that a metric is not increasing by a quantum-classical (QC) channel,
and coincides with classical Fisher information restricted to classical
probability distributions. Then, this metric should be no smaller than SLD
Fisher metric: if one apply the optimal measurement $M$,
\[
J_{\rho_{\theta}}^{S}=J_{M\left(  \rho_{\theta}\right)  }=g_{M\left(
\rho_{\theta}\right)  },
\]
where the second identity is the assumption of normalization: therefore, due
to the monotonicity by the measurement $M$,
\[
g_{\rho_{\theta}}\geq g_{M\left(  \rho_{\theta}\right)  }=J_{\rho_{\theta}%
}^{S}.
\]

Similarly, assume that a metric is not increasing by a classical-quantum (CQ)
channel and coincides with classical Fisher information for probability
distributions. Then, the metric should be no larger than RLD Fisher metric: an
optimal tangent reverse estimation $\left(  \Phi,\left\{  p_{\theta
},\mathrm{d}p_{\theta}/\mathrm{d}\theta\right\}  \right)  \,$of the
$\{\rho_{\theta},\mathrm{d}\rho_{\theta}/\mathrm{d}\theta\}$ satisfies
\[
J_{\rho_{\theta}}^{R}=J_{p_{\theta}}=g_{p_{\theta}}\,,\,\,\,\,
\]
where the second identity is due to the assumption of normalization:
therefore, due to monotonicity by\ the CQ map $\Phi$,
\[
g_{\rho_{\theta}}\leq g_{p_{\theta}}=J_{\rho_{\theta}}^{R}.
\]

Here, we have not assumed that the metric is Riemannian, or induced from an
inner product in the tangent space, different from the argument in
\cite{Petz}. Also, we have only assumed monotonicity by QC and CQ maps:

\begin{theorem}
Assume that a (not necessarily Riemannian) metric $g$ coincide with classical
Fisher information in the space of classical probability distributions. Then,
if $g$ is monotone decreasing by a QC map, $g$ is no smaller than SLD Fisher
metric. If $\ g$ is monotone decreasing by a CQ map, $g$ is no larger than RLD
Fisher metric.
\end{theorem}

\begin{example}
(Petz metrics) \ In \cite{Petz}, Petz had shown any monotone Riemannian metric
can be written as
\[
g_{\rho_{\theta}}^{f}\left(  \partial_{i}\rho_{\theta},\partial_{j}%
\rho_{\theta}\right)  :=\mathrm{tr}\,\partial_{i}\rho_{\theta}\,\left\{
\mathbf{R}_{\rho_{\theta}}f\left(  \mathbf{L}_{\rho_{\theta}}\mathbf{R}%
_{\rho_{\theta}}^{-1}\right)  \right\}  ^{-1}\partial_{j}\rho_{\theta},
\]
where \ $\mathbf{L}_{\rho_{\theta}}$ and $\mathbf{R}_{\rho_{\theta}}$ are map
form $\mathfrak{B}\left(  \mathcal{H}\right)  $ to $\mathfrak{B}\left(
\mathcal{H}\right)  $ with
\[
\mathbf{L}_{\rho_{\theta}}\left(  A\right)  =\rho_{\theta}A,\,\,\mathbf{R}%
_{\rho_{\theta}}\left(  A\right)  =A\rho_{\theta},
\]
and $f$ is an operator monotone function with
\[
f\left(  x\right)  =xf\left(  x^{-1}\right)  ,\,\,\,\,f\left(  1\right)  =1.
\]
For RLD and SLD metric, $f\left(  x\right)  =\frac{2x}{x+1}$ and $f\left(
x\right)  =\frac{x+1}{2}$, respectively. If $f\left(  x\right)  =\frac
{x-1}{\log x}$,
\[
g_{\rho_{\theta}}^{f}\left(  \partial_{i}\rho_{\theta},\partial_{j}%
\rho_{\theta}\right)  =\mathrm{tr}\,\partial_{i}\rho_{\theta}\partial_{j}%
\ln\rho_{\theta}:=J_{\theta,i,j}^{B}\,,
\]
which is called Bogoljubov-Kubo-Mori (BKM) metric. It had been known that
\[
J_{\theta}^{S}\leq J_{\theta}^{B}\leq J_{\theta}^{R}.
\]
Also, \cite{HasegawaPetz}
\[
f\left(  x\right)  =f_{\alpha}\left(  x\right)  =\left(  1-\frac{\alpha^{2}%
}{4}\right)  \frac{\left(  x-1\right)  ^{2}}{\left(  x^{\frac{1-\alpha}{2}%
}-1\right)  \left(  x^{\frac{1+\alpha}{2}}-1\right)  }\,\,\,\left(  \left\vert
\alpha\right\vert \leq3\right)
\]
is operator monotone for $\left\vert \alpha\right\vert \leq3$, and
corresponding metric will be denoted by $J_{\theta}^{\alpha}$, hereafter. It
holds that
\[
J_{\theta}^{3}=J_{\theta}^{-3}=J_{\theta}^{R},\quad J_{\theta}^{1}=J_{\theta
}^{-1}=J_{\theta}^{B}.
\]
Hence, $J_{\theta}^{\alpha}$ ($1\leq\alpha\leq3$) `interpolates' between
$J_{\theta}^{B}$ and $J_{\theta}^{R}$. In addition, \cite{HasegawaPetz} shown
\begin{equation}
J_{\theta}^{S}\leq J_{\theta}^{0}\leq J_{\theta}^{B}\text{.} \label{J0>JS}%
\end{equation}
\ 
\end{example}

\section{Non-asymptotic scenario}

\subsection{Parallel family of states \ }

A family $\left\{  \rho_{\theta}\right\}  $ is said to be
\textit{RLD-parallel} if and only if:%
\[
\rho_{\theta}=NM_{\theta}N^{\dagger},\,\,
\]
and%
\begin{align*}
&  M_{\theta}=\mathrm{diag}\,\left(  p_{\theta}\left(  1\right)  ,p_{\theta
}\left(  2\right)  ,\cdots,p_{\theta}\left(  r\right)  \right)  ,\\
N  &  =\left[  \left\vert \phi_{1}\right\rangle ,\left\vert \phi
_{2}\right\rangle ,\cdots,\left\vert \phi_{r}\right\rangle \right]
\end{align*}
where $\left\{  \left\vert \phi_{1}\right\rangle ,\,\cdots,\left\vert \phi
_{r}\right\rangle \right\}  $ is a linearly independent, normalized, but not
necessarily orthogonal system of state vectors. This condition is equivalent
to
\[
\rho_{\theta}=\sum_{x=1}^{r}p_{\theta}\left(  x\right)  \left\vert \phi
_{x}\right\rangle \left\langle \phi_{x}\right\vert .
\]
Its operational meaning is as follows. Observe that $\left(  \Phi,\left\{
p_{\theta}\right\}  \right)  $ is the reverse estimation of $\left\{
\rho_{\theta}\right\}  $, with $\Phi\left(  \delta_{x}\right)  =\left\vert
\phi_{x}\right\rangle \left\langle \phi_{x}\right\vert $. The Fisher
information $J_{\theta}$ of $\left\{  p_{\theta}\right\}  $ is easily computed
by observing
\[
L_{\theta,i}^{R}=N\mathrm{diag}\,\left(  \partial_{i}\ln p_{\theta}\left(
1\right)  ,\cdots,\partial_{i}\ln p_{\theta}\left(  r\right)  \right)
N^{-1},
\]
(where $N^{-1}$ is the Moor-Penrose generalized inverse) and we obtain
\begin{equation}
J_{\theta}=J_{\theta}^{R}. \label{parallel-JR}%
\end{equation}
Hence this reverse estimation achieves the lowerbound suggested by
Theorem\thinspace\ref{th:r-rld-bound} at any $\theta$.

Hereafter, let
\[
p_{t}^{\left(  m\right)  }:=t\,p+\left(  1-t\right)  q,\,\,\rho_{t}^{\left(
m\right)  }:=t\,\rho+\left(  1-t\right)  \sigma.
\]

\begin{proposition}
\label{prop:two-point-parallel}For any $\rho$ and $\sigma$ with $\mathrm{supp}%
\,\rho=\mathrm{supp}\,\sigma$, there is an RLD-parallel manifold containing
$\rho$, $\sigma$, and $\rho_{t}^{\left(  m\right)  }$, for any $0\leq t\leq1$.
\end{proposition}

\begin{proof}
Let $P$ be the projection onto $\mathrm{supp}\,\rho=\mathrm{supp}\,\sigma$.
Let $U$ be a unitary matrix such that $PU\left(  1-P\right)  =0$ and
\[
\sqrt{\rho}U\sqrt{\sigma}=\sqrt{\sigma}U^{\dagger}\sqrt{\rho},
\]
or equivalently
\[
\sqrt{\sigma^{-1}}\sqrt{\rho}U=U^{\dagger}\sqrt{\rho}\sqrt{\sigma^{-1}},
\]
where $\rho^{-1}$ and $\sigma^{-1}$ are generalized inverse. Such $U$ is found
out using the polar decomposition of $\sqrt{\sigma^{-1}}\sqrt{\rho}$. By
Proposition\thinspace\ref{th:A=BL}, there is $X$ with
\[
\sqrt{\rho}U=\sqrt{\sigma}X,\quad\,X=X^{\dagger}.
\]
Let $VDV^{\dagger}$ be diagonalization of $X$, and we obtain
\[
\sqrt{\rho}UV=\sqrt{\sigma}VD.
\]
Divide $x$th column vector of $\sqrt{\sigma}V$ by its magnitude and denote the
product by $\left\vert \phi_{x}\right\rangle $. Then letting $N:=\left[
\left\vert \phi_{1}\right\rangle ,\left\vert \phi_{2}\right\rangle
,\cdots,\left\vert \phi_{r}\right\rangle \right]  $, we have
\begin{align*}
\rho &  =N\mathrm{diag}\,\left(  p\left(  1\right)  ,\cdots,p\left(  r\right)
\right)  N^{\dagger},\,\sigma=N\mathrm{diag}\,\left(  q\left(  1\right)
,\cdots,q\left(  r\right)  \right)  N^{\dagger},\\
\rho_{t}^{\left(  m\right)  }  &  =N\mathrm{diag}\,\left(  p_{t}^{\left(
m\right)  }\left(  1\right)  ,\cdots,p_{t}^{\left(  m\right)  }\left(
r\right)  \right)  N^{\dagger},
\end{align*}
for some $p\left(  1\right)  $, $\cdots$, $p\left(  r\right)  $, and $q\left(
1\right)  $, $\cdots$, $q\left(  r\right)  $, and the assertion is proved.
\end{proof}

\subsection{Reverse test}

\label{subsec:reverse-test}Consider test of the hypothesis `the given state is
$\rho$' against the alternative hypothesis `the given state is $\sigma$'. (
Hereafter, such test is referred to as " test `$\rho$ vs. $\sigma$' ".)
Suppose we are given many copies of the unknown states, and the error $a_{n}$
of the first kind, or the probability of rejecting $\rho$ while $\rho$ is the
true state, vanishes as $n\rightarrow\infty$. Then in maximizing the exponent
of the error $\beta_{n}$ of the second kind, or the probability of rejecting
$\sigma$ while $\sigma$ is the true state, the key step is optimization of QC
map (measurement) $M$ to maximize the relative entropy $\mathrm{D}\left(
M\left(  \rho^{\otimes n}\right)  ||M\left(  \sigma^{\otimes n}\right)
\right)  $.

We consider \textit{reverse test}, or the inverse process of (the single copy
version of ) the above. Given a pair $\left\{  \rho,\sigma\right\}  $ of
states, let $\Phi$ be CQ map with
\[
\Phi\left(  p\right)  =\rho,\,\Phi\left(  q\right)  =\sigma,
\]
where $\left\{  p,q\right\}  $ is a pair of probability distributions. \ A
pair $\left(  \Phi,\left\{  p.q\right\}  \right)  $ is called a reverse test
of $\left\{  \rho,\sigma\right\}  $. (In the terminology of statistical
decision theory, $\left\{  \rho,\sigma\right\}  $ is deficient relative to
$\left\{  p,q\right\}  $.) Our task is to minimize $\mathrm{D}\left(
p||q\right)  $ for all reverse tests.

To find the optimal reverse test, the following lemma plays a key role :

\begin{lemma}
\label{lem:D-int}(\cite{AmariNagaoka}, Chapter 3, Section 3.5)
\[
\mathrm{D}\left(  p||q\right)  =\int_{0}^{1}\int_{0}^{t}J_{p_{s}^{\left(
m\right)  }}\mathrm{d}s\mathrm{d}t\,.
\]

\end{lemma}

Let $\left(  \Phi,\left\{  p.q\right\}  \right)  $ be a reverse test of
$\left\{  \rho,\sigma\right\}  $. Then
\[
\Phi\left(  p_{t}^{\left(  m\right)  }\right)  =\rho_{t}^{\left(  m\right)  }%
\]
holds, and$\left(  \Phi,\left\{  p_{t}^{\left(  m\right)  }\right\}  \right)
$ is a reverse estimation of $\left\{  \rho_{t}^{\left(  m\right)  }\right\}
$. Therefore, by Lemma\thinspace\ref{lem:D-int}\ and Theorem\thinspace
\ref{th:r-rld-bound}, we have
\[
\mathrm{D}\left(  p||q\right)  \geq\int_{0}^{1}\int_{0}^{t}J_{\rho
_{s}^{\left(  m\right)  }}^{R}\mathrm{d}s\mathrm{d}t\,.
\]
Also, by \ref{prop:two-point-parallel}, there is a parallel family which
contains $\left\{  \rho_{t}^{\left(  m\right)  }\right\}  $. Therefore,
\[
\rho_{\theta}=N\mathrm{diag}\,\left(  p_{t}^{\left(  m\right)  }\left(
1\right)  ,\cdots,p_{t}^{\left(  m\right)  }\left(  r\right)  \right)
N^{\dagger},
\]
holds for some $N=[\left\vert \phi_{1}\right\rangle ,\cdots,\left\vert
\phi_{r}\right\rangle ]$. Hence, by (\ref{parallel-JR}), the reverse
estimation $\left(  \Phi_{0},\left\{  \,p_{\theta}\right\}  \right)  $, where
$\Phi_{0}\left(  \delta_{x}\right)  =\left\vert \phi_{x}\right\rangle
\left\langle \phi_{x}\right\vert $, achieves \
\[
J_{\rho_{t}^{\left(  m\right)  }}^{R}=J_{p_{t}^{\left(  m\right)  }},\,\forall
t,
\]
and $\Phi_{0}\left(  p\right)  =\rho$, $\Phi_{0}\left(  q\right)  =\sigma$.
Therefore, the reverse test $\left(  \Phi_{0},\left\{  \,p,q\right\}  \right)
$ achieves%

\[
\int_{0}^{1}\int_{0}^{t}J_{\rho_{t}^{\left(  m\right)  }}^{R}\mathrm{d}%
s\mathrm{d}t=\int_{0}^{1}\int_{0}^{t}J_{p_{s}^{\left(  m\right)  }}%
\mathrm{d}s\mathrm{d}t=\sum_{x=1}^{r}p\left(  x\right)  \ln\frac{p\left(
x\right)  }{q\left(  x\right)  }\text{,}%
\]
and hence is optimal. The right most side integral is computed in
\cite{Hayashi:2005}, although the detail is not described. Here we show a way
to verify that the left most side equals $\mathrm{D}^{R}\left(  \rho
||\sigma\right)  $, as in \cite{Matsumoto:2005}. \ Observe that there is a
$r\times r$ unitary matrix $U$ with
\begin{align*}
\rho^{\frac{1}{2}}  &  =UD_{0}N^{\dagger}=ND_{0}U^{\dagger},\\
\sigma^{\frac{1}{2}}  &  =UD_{1}N^{\dagger}=ND_{1}U^{\dagger},\\
D_{t}  &  :=\mathrm{diag}\,\left(  \sqrt{p_{t}^{\left(  m\right)  }\left(
1\right)  },\cdots,\sqrt{p_{t}\left(  r\right)  }\right)  .
\end{align*}
Therefore,\noindent\
\begin{align*}
&  \mathrm{tr}\,\rho\ln\rho^{\frac{1}{2}}\sigma^{-1}\rho^{\frac{1}{2}}\\
&  =\!\!{}\mathrm{tr}\left[  UD_{0}N^{\dagger}ND_{0}U^{\dagger}\ln\left\{
UD_{0}N^{\dagger}\left(  UD_{1}N^{\dagger}\right)  ^{-1}\left(  ND_{1}%
U^{\dagger}\right)  ^{-1}ND_{0}U^{\dagger}\right\}  \right] \\
&  =\mathrm{tr}\,N^{\dagger}N\mathrm{diag}\,\left(  p\left(  1\right)
\ln\frac{p\left(  1\right)  }{q\left(  1\right)  },\cdots,p\left(  r\right)
\ln\frac{p\left(  r\right)  }{q\left(  r\right)  }\right) \\
&  =\sum_{x}\left\Vert \phi_{x}\right\Vert ^{2}p\left(  x\right)  \ln
\frac{p\left(  x\right)  }{q\left(  x\right)  }=\sum_{x}p\left(  x\right)
\ln\frac{p\left(  x\right)  }{q\left(  x\right)  }.
\end{align*}
Thus we obtain:

\textbf{Theorem 2.4} \ \
\[
\min\mathrm{D}\left(  p||q\right)  =\mathrm{D}^{R}\left(  \rho||\sigma\right)
,
\]
\textit{where minimization is taken for over all reverse tests }$\left(
\Phi,\left\{  p,q\right\}  \right)  $\textit{\ of }$\left\{  \rho
,\sigma\right\}  $\textit{.}

\subsection{Monotone relative entropy}

An example of the quantity satisfying (M) and (N) is $\mathrm{D}(\rho
||\sigma)=$ $\mathrm{Tr}\rho\left(  \ln\rho-\ln\sigma\right)  $. By
Theorem\thinspace\ref{th:DM<DQ<DR}, we obtain another proof of the inequality
shown in \cite{HiaiPetz},
\[
\mathrm{D}(\rho||\sigma)\leq\mathrm{D}^{R}(\rho||\sigma).
\]
Another example is
\[
\mathrm{D}^{g}(\rho||\sigma):=\int_{0}^{1}\int_{0}^{t}g_{\rho_{s}^{\left(
m\right)  }}\mathrm{d}s\mathrm{d}t,
\]
where $g$ is any properly normalized monotone metric. Note $\mathrm{D}%
^{R}=\mathrm{D}^{J^{R}}$. Also, it is known \cite{Nagaoka:1991}%
\cite{Hayashi:2005}\ that
\begin{equation}
\mathrm{D}^{J^{B}}(\rho||\sigma)=\mathrm{D}(\rho||\sigma). \label{Db=D}%
\end{equation}
Due to Lemma\thinspace\ref{lem:D-int}, $\mathrm{D}^{g}(p||q)=\mathrm{D}(p||q)$
for all probability distributions $p$, $q$. Also, since
\[
\Lambda\left(  \rho_{t}^{\left(  m\right)  }\right)  =\left(  1-t\right)
\Lambda\left(  \rho\right)  +t\,\Lambda\left(  \sigma\right)  ,
\]
$\mathrm{D}^{g}(\rho||\sigma)$ is monotone decreasing by application of CPTP
maps:
\begin{align*}
\mathrm{D}^{g}(\Lambda\left(  \rho\right)  ||\Lambda\left(  \sigma\right)  )
&  =\int_{0}^{1}\int_{0}^{t}g_{\Lambda\left(  \rho_{s}^{\left(  m\right)
}\right)  }\mathrm{d}s\mathrm{d}t\\
&  \leq\int_{0}^{1}\int_{0}^{t}g_{\rho_{s}^{\left(  m\right)  }}%
\mathrm{d}s\mathrm{d}t=\mathrm{D}^{g}(\rho||\sigma).
\end{align*}

\begin{corollary}%
\[
\lim_{n\rightarrow\infty}\frac{1}{n}\mathrm{D}^{J^{S}}(\rho^{\otimes
n}||\sigma^{\otimes n})=\mathrm{D}(\rho||\sigma),\quad\,\lim_{n\rightarrow
\infty}\frac{1}{n}\mathrm{D}^{J^{0}}(\rho^{\otimes n}||\sigma^{\otimes
n})=\mathrm{D}(\rho||\sigma).\quad
\]

\end{corollary}

\begin{proof}
Since both of
\[
\varlimsup_{n\rightarrow\infty}\frac{1}{n}\mathrm{D}^{J^{S}}(\rho^{\otimes
n}||\sigma^{\otimes n}),\,\,\varliminf_{n\rightarrow\infty}\frac{1}%
{n}\mathrm{D}^{J^{S}}(\rho^{\otimes n}||\sigma^{\otimes n})
\]
satisfy (M), (N), and (A), Theorem\thinspace\ref{th:D<DQ<DR} implies
\begin{align*}
&  \varlimsup_{n\rightarrow\infty}\frac{1}{n}\mathrm{D}^{J^{S}}(\rho^{\otimes
n}||\sigma^{\otimes n})\\
&  \geq\varliminf_{n\rightarrow\infty}\frac{1}{n}\mathrm{D}^{J^{S}}%
(\rho^{\otimes n}||\sigma^{\otimes n})\geq\mathrm{D}(\rho||\sigma).
\end{align*}
On the other hand, since $\mathrm{D}(\rho||\sigma)=\mathrm{D}^{J^{B}}%
(\rho||\sigma)$ and $J^{B}\geq J^{S}$,
\[
\mathrm{D}(\rho||\sigma)=\frac{1}{n}\mathrm{D}(\rho^{\otimes n}||\sigma
^{\otimes n})\geq\frac{1}{n}\mathrm{D}^{J^{S}}(\rho^{\otimes n}||\sigma
^{\otimes n}).
\]
After all,
\begin{align*}
\mathrm{D}(\rho||\sigma)  &  \geq\varlimsup_{n\rightarrow\infty}\frac{1}%
{n}\mathrm{D}^{J^{S}}(\rho^{\otimes n}||\sigma^{\otimes n})\\
&  \geq\varliminf_{n\rightarrow\infty}\frac{1}{n}\mathrm{D}^{J^{S}}%
(\rho^{\otimes n}||\sigma^{\otimes n})\geq\mathrm{D}(\rho||\sigma).
\end{align*}
Due to (\ref{J0>JS}), the second identity follows from the first.
\end{proof}

\begin{theorem}%
\[
\lambda\mathrm{D}^{R}(\rho_{0}||\sigma_{0})+\left(  1-\lambda\right)
\mathrm{D}^{R}(\rho_{1}||\sigma_{1})\geq\mathrm{D}^{R}(\lambda\rho_{0}+\left(
1-\lambda\right)  \rho_{1}||\lambda\sigma_{0}+\left(  1-\lambda\right)
\sigma_{1}).
\]

\end{theorem}

\begin{proof}
Let $\left(  \Phi_{y},\left\{  p_{y},q_{y}\right\}  \right)  $ be an optimal
reverse test of $\left\{  \rho_{y},\sigma_{y}\right\}  $ ($y=0,1$). Define
$\tilde{p}_{y_{0}}\left(  x,y\right)  :=p_{y}\left(  x\right)  \delta_{y_{0}%
}\left(  y\right)  $, $\tilde{q}_{y_{0}}\left(  x,y\right)  :=q_{y}\left(
x\right)  \delta_{y_{0}}\left(  y\right)  $, \ and $\tilde{\Phi}\left(
\delta_{\left(  x,y\right)  }\right)  =\Phi_{y}\left(  x\right)  $. Then we
have
\begin{align*}
\tilde{\Phi}\left(  \tilde{p}_{y_{0}}\right)   &  =\sum_{x,y}p_{y}\left(
x\right)  \delta_{y_{0}}\left(  y\right)  \Phi_{y_{0}}\left(  x\right)
=\rho_{y_{0}},\\
\,\tilde{\Phi}\left(  \tilde{q}_{y_{0}}\right)   &  =\sigma_{y_{0}},\\
\mathrm{D}\left(  \tilde{p}_{y_{0}}||\tilde{q}_{y_{0}}\right)   &  =\sum
_{x,y}p_{y}\left(  x\right)  \delta_{y_{0}}\left(  y\right)  \ln\frac
{p_{y}\left(  x\right)  \delta_{y_{0}}\left(  y\right)  }{q_{y}\left(
x\right)  \delta_{y_{0}}\left(  y\right)  }\\
&  =\mathrm{D}\left(  p_{y_{0}}||q_{y_{0}}\right)  =\mathrm{D}^{R}\left(
\rho_{y_{0}}||\sigma_{y_{0}}\right)  .
\end{align*}
Therefore,
\begin{align*}
&  \lambda\mathrm{D}^{R}(\rho_{0}||\sigma_{0})+\left(  1-\lambda\right)
\mathrm{D}^{R}(\rho_{1}||\sigma_{1})=\lambda\mathrm{D}(\tilde{p}_{0}%
||\tilde{q}_{0})+\left(  1-\lambda\right)  \mathrm{D}(\tilde{p}_{1}||\tilde
{q}_{1})\\
&  \geq\mathrm{D}(\lambda\tilde{p}_{0}+\left(  1-\lambda\right)  \tilde{p}%
_{1}||\lambda\tilde{q}_{0}+\left(  1-\lambda\right)  \tilde{q}_{1})\\
&  \geq\mathrm{D}^{R}(\tilde{\Phi}\left(  \lambda\tilde{p}_{0}+\left(
1-\lambda\right)  \tilde{p}_{1}\right)  ||\tilde{\Phi}\left(  \lambda\tilde
{q}_{0}+\left(  1-\lambda\right)  \tilde{q}_{1}\right)  )\\
&  =\mathrm{D}^{R}(\lambda\rho_{0}+\left(  1-\lambda\right)  \rho_{1}%
||\lambda\sigma_{0}+\left(  1-\lambda\right)  \tilde{\sigma}_{1}).
\end{align*}

\end{proof}

\section{Asymptotic scenario}

\subsection{Asymptotic reverse test}

The result of the test `$\rho^{n}$ vs. $\sigma^{n}$\thinspace' is binary, that
is, accept $\rho^{n}$ or $\sigma^{n}$ . Hence, a natural inverse problem would
be generation of $\left\{  \rho^{n},\sigma^{n}\right\}  $ from the probability
distributions $\left\{  p^{n},q^{n}\right\}  $ over binary set $\left\{
0,1\right\}  $. Let us define an \textit{asymptotic reverse test}, or
\textit{\ }a pair $\left(  \Phi^{n},\left\{  p^{n},q^{n}\right\}  \right)  $
with
\begin{align}
\lim_{n\rightarrow\infty}\left\Vert \Phi^{n}\left(  p^{n}\right)  -\rho
^{n}\right\Vert _{1}  &  =0,\,\Phi^{n}\left(  q^{n}\right)  =\sigma
^{n},\nonumber\\
\,\,\lim_{n\rightarrow\infty}p^{n}\left(  0\right)   &  =\lim_{n\rightarrow
\infty}q^{n}\left(  1\right)  =1, \label{e-reverse-test}%
\end{align}
and discuss the infimum of
\[
\varliminf_{n\rightarrow\infty}\frac{-1}{n}\ln q^{n}\left(  0\right)  .
\]

To describe the infimum, we need the following object:%
\begin{align*}
&  \mathrm{D}_{\max}^{\infty}\left(  \{\rho^{n}\}||\{\sigma^{n}\}\right) \\
&  :=\inf\left\{  a\,;\text{ }\tilde{\rho}^{n}\leq e^{na}\,\sigma^{n}%
,\lim_{n\rightarrow\infty}\left\Vert \tilde{\rho}^{n}-\rho^{n}\right\Vert
_{1}=0\right\}  .
\end{align*}
The following proposition is trivial.

\begin{proposition}
$\mathrm{D}_{\max}^{\infty}$ is monotone decreasing by application of a CPTP
map $\Lambda$,%
\begin{equation}
\mathrm{D}_{\max}^{\infty}\left(  \{\Lambda\left(  \rho^{n}\right)
\}||\{\Lambda\left(  \sigma^{n}\right)  \}\right)  \leq\mathrm{D}_{\max
}^{\infty}\left(  \{\rho^{n}\}||\{\sigma^{n}\}\right)  , \label{monotone-Dmin}%
\end{equation}
and asymptotically continuous about the first argument,
\begin{align}
\lim_{n\rightarrow\infty}\left\Vert \tilde{\rho}^{n}-\rho^{n}\right\Vert =  &
0\,\Longrightarrow\nonumber\\
\mathrm{D}_{\max}^{\infty}\left(  \{\tilde{\rho}^{n}\}||\{\sigma^{n}\}\right)
&  =\mathrm{D}_{\max}^{\infty}\left(  \{\rho^{n}\}||\{\sigma^{n}\}\right)  .
\label{cont-Dmin}%
\end{align}

\end{proposition}

\begin{theorem}
\label{th:asym-reverse-test}%
\[
\inf\varliminf_{n\rightarrow\infty}\frac{-1}{n}\ln q^{n}\left(  0\right)
=\mathrm{D}_{\max}^{\infty}\left(  \{\rho^{n}\}||\{\sigma^{n}\}\right)
\]
where $\inf$ is taken over all the asymptotic reverse test.
\end{theorem}

\begin{proof}
First, we show `$\leq$'. By definition of $\mathrm{D}_{\max}^{\infty}$, for
any $c>0$, it is possible to define $\Phi^{n}\left(  \delta_{0}\right)  $ so
that
\begin{align*}
&  \lim_{n\rightarrow\infty}\left\Vert \Phi^{n}\left(  \delta_{0}\right)
-\rho^{n}\right\Vert _{1}=0,\\
\Phi^{n}\left(  \delta_{0}\right)   &  \leq\sigma^{n}\exp\left\{  n\left(
\mathrm{D}_{\max}^{\infty}\left(  \{\rho^{n}\}||\{\sigma^{n}\}\right)
+c\right)  \right\}  .
\end{align*}
hold. Then, letting
\begin{align*}
\lim_{n\rightarrow\infty}p^{n}\left(  0\right)   &  =1,\\
q^{n}\left(  0\right)   &  =\exp\left\{  -n\left(  \mathrm{D}_{\max}^{\infty
}\left(  \{\rho^{n}\}||\{\sigma^{n}\}\right)  +c\right)  \right\}  ,
\end{align*}
we have
\begin{align*}
\lim_{n\rightarrow\infty}\left\Vert \Phi^{n}\left(  p^{n}\right)  -\rho
^{n}\right\Vert _{1}  &  =\lim_{n\rightarrow\infty}\left\Vert \Phi^{n}\left(
\delta_{0}\right)  -\rho^{n}\right\Vert _{1}=0,\\
\sigma^{n}-q^{n}\left(  0\right)  \Phi^{n}\left(  \delta_{0}\right)   &
\geq0.
\end{align*}
Therefore, it is possible to define $\Phi^{n}\left(  \delta_{1}\right)  $ so
that%
\[
\Phi^{n}\left(  q^{n}\right)  =q^{n}\left(  0\right)  \Phi^{n}\left(
\delta_{0}\right)  +q^{n}\left(  1\right)  \Phi^{n}\left(  \delta_{1}\right)
=\sigma^{n}%
\]
holds. To sum up, a sequence of reverse test $\left(  \Phi^{n},\left\{
p^{n},q^{n}\right\}  \right)  $ satisfies the requirement
(\ref{e-reverse-test}), and satisfies
\[
\varliminf_{n\rightarrow\infty}\frac{-1}{n}\ln q^{n}\left(  0\right)
=\mathrm{D}_{\max}^{\infty}\left(  \{\rho^{n}\}||\{\sigma^{n}\}\right)  +c.
\]
Since this composition is possible for any $c>0$, we have `$\leq$'.

Second, we prove `$\geq$'. Observe, due to (\ref{e-reverse-test}),
\[
\inf\varliminf_{n\rightarrow\infty}\frac{-1}{n}\ln q^{n}\left(  0\right)
=\mathrm{D}_{\max}^{\infty}\left(  \left\{  p^{n}\right\}  ||\left\{
q^{n}\right\}  \right)  .
\]
Therefore, by monotonicity (\ref{monotone-Dmin}) of $\mathrm{D}_{\max}%
^{\infty}$,%
\begin{align*}
\inf\varliminf_{n\rightarrow\infty}\frac{-1}{n}\ln q^{n}\left(  0\right)   &
\geq\mathrm{D}_{\max}^{\infty}\left(  \Phi^{n}\left(  p^{n}\right)  ||\Phi
^{n}\left(  q^{n}\right)  \right) \\
&  =\mathrm{D}_{\max}^{\infty}\left(  \Phi^{n}\left(  p^{n}\right)
||\sigma^{n}\right)  .
\end{align*}
This, by asymptotic continuity (\ref{cont-Dmin}) and (\ref{e-reverse-test}),
leads to `$\geq$'.
\end{proof}

A converse statement of Stein's lemma can be maid using $\mathrm{D}_{\max
}^{\infty}$, indicating asymptotic reverse test is a natural inverse problem
of test.

\begin{corollary}
\label{cor:stein-converse}(converse statement of Stein's lemma)%
\[
\mathrm{D}_{\max}^{\infty}\left(  \{\rho^{n}\}||\{\sigma^{n}\}\right)
\geq\sup\left\{  \varliminf_{n\rightarrow\infty}\,\frac{-1}{n}\log
\mathrm{tr}\,P^{n}\sigma^{n};\,\varliminf_{n\rightarrow0}\mathrm{tr}\,\rho
^{n}P^{n}>0,\,0\leq P_{n}\leq\mathbf{1}\right\}  .
\]

\end{corollary}

\begin{proof}
Consider $\left(  \Phi^{n},\left\{  p^{n},q^{n}\right\}  \right)  $ with
(\ref{e-reverse-test}), and let $\tilde{p}^{n}$ and $\tilde{q}^{n}$ be binary
distributions with
\[
\tilde{p}^{n}\left(  0\right)  :=\mathrm{tr}\,P^{n}\Phi^{n}\left(
p^{n}\right)  ,\,\,\tilde{q}^{n}\left(  0\right)  :=\mathrm{tr}\,P^{n}\Phi
^{n}\left(  q^{n}\right)  ,
\]
and let $P^{n}$ be a POVM element with $\varliminf_{n\rightarrow0}%
\mathrm{tr}\,\rho^{n}P^{n}>0$. Then,
\begin{equation}
\varliminf_{n\rightarrow0}\tilde{p}^{n}\left(  0\right)  =\varliminf
_{n\rightarrow0}\mathrm{tr}\,P^{n}\Phi^{n}\left(  p^{n}\right)  \geq
\varliminf_{n\rightarrow0}\left\{  \mathrm{tr}\,P^{n}\rho^{n}-\left\Vert
\rho^{n}-\Phi^{n}\left(  p^{n}\right)  \right\Vert _{1}\right\}  >0.
\label{ptilde<1}%
\end{equation}

Observe that the composition of the map $\Phi^{n}$ and the measurement
$\left\{  P^{n},\mathbf{1}-P^{n}\right\}  $ is a CPTP map, or a Markov map
from binary distributions onto themselves. Hence, it should be written as
\begin{align*}
\tilde{p}^{n}\left(  0\right)   &  =a_{00}^{n}\,p^{n}\left(  0\right)
+a_{01}^{n}\,p^{n}\left(  1\right)  ,\\
\tilde{q}^{n}\left(  0\right)   &  =a_{00}^{n}\,q^{n}\left(  0\right)
+a_{01}^{n}\,q^{n}\left(  1\right)  .
\end{align*}
By (\ref{ptilde<1}), we should have $\ \varliminf_{n\rightarrow\infty}%
a_{00}^{n}>0$ or $\varliminf_{n\rightarrow\infty}a_{01}^{n}>0\,$. \ If the
former is true,
\[
\varliminf_{n\rightarrow\infty}\frac{-1}{n}\log q^{n}\left(  0\right)
\geq\varliminf_{n\rightarrow\infty}\frac{-1}{n}\log\tilde{q}^{n}\left(
0\right)  -\varlimsup_{n\rightarrow\infty}\frac{-1}{n}\log a_{00}%
^{n}=\varliminf_{n\rightarrow\infty}\frac{-1}{n}\log\tilde{q}^{n}\left(
0\right)  .
\]
If the latter is true, due to $\lim_{n\rightarrow\infty}q^{n}\left(  1\right)
=1$, we have
\[
\varliminf_{n\rightarrow\infty}\frac{-1}{n}\log\tilde{q}^{n}\left(  0\right)
=0.
\]
In either case, we have
\[
\varliminf_{n\rightarrow\infty}\frac{-1}{n}\log q^{n}\left(  0\right)
\geq\varliminf_{n\rightarrow\infty}\frac{-1}{n}\log\tilde{q}^{n}\left(
0\right)  =\varliminf_{n\rightarrow\infty}\frac{-1}{n}\log\mathrm{tr}%
\,P_{n}\sigma^{n}.
\]
Also, Theorem\thinspace\ref{th:asym-reverse-test}\ implies that there is
$\left(  \Phi^{n},\left\{  p^{n},q^{n}\right\}  \right)  $ with
\[
\varliminf_{n\rightarrow\infty}\frac{-1}{n}\log q^{n}\left(  0\right)
\leq\mathrm{D}_{\max}^{\infty}\left(  \{\rho^{n}\}||\{\sigma^{n}\}\right)
+c,
\]
for any $c>0$. Therefore, we have to have the converse statement.
\end{proof}

\subsection{Relations between $\mathrm{D}_{\max}^{\infty}$ , $\overline
{\mathrm{D}}$ and $\mathrm{D}$}

Nagaoka\thinspace\cite{Nagaoka:1999} defined the following quantity to analyze
quantum hypothesis test :
\[
\overline{\mathrm{D}}\left(  \{\rho^{n}\}||\{\sigma^{n}\}\right)
:=\inf\left\{  a\,;\lim_{n\rightarrow\infty}\,\mathrm{tr}\,\rho^{n}\left\{
\rho^{n}-e^{na}\sigma^{n}\leq0\right\}  =1\right\}  ,
\]
where $\left\{  \rho^{n}-e^{na}\sigma^{n}\leq0\right\}  $ is the projector
onto the non-positive eigenspace of $\rho^{n}-$ $e^{na}\,\sigma^{n}$.

\begin{theorem}
\label{th:NH-converse}\cite{Nagaoka:1999}\cite{NagaokaHayashi} $\overline
{\mathrm{D}}\left(  \{\rho_{n}\}||\{\sigma_{n}\}\right)  $ characterizes
efficiency of the test `$\rho^{n}$ vs. $\sigma^{n}\,$' as follows.%
\[
\overline{\mathrm{D}}\left(  \{\rho^{n}\}||\{\sigma^{n}\}\right)
=\inf\left\{  a;\forall\left\{  P^{n}\right\}  \,\varliminf_{n\rightarrow
\infty}\frac{-1}{n}\ln\mathrm{tr}\,P^{n}\sigma^{n}\geq a\Rightarrow
\lim_{n\rightarrow\infty}\mathrm{tr}\,P^{n}\rho^{n}=0\right\}  .
\]

\end{theorem}

\begin{lemma}
\label{lem:datta}(Datta\thinspace\cite{Datta}) If
\[
\varepsilon:=1-\mathrm{tr}\,\rho^{n}\left\{  \rho^{n}-e^{na}\sigma^{n}%
\leq0\right\}  ,
\]
there is a positive operator $A^{n}$ with
\[
\left\Vert A^{n}-\rho^{n}\right\Vert _{1}\leq\sqrt{8\varepsilon}%
\,,\,\,A^{n}\leq e^{na}\sigma^{n}.
\]

\end{lemma}

Let $A^{n}$ as of Lemma\thinspace\ref{lem:datta}, and $\tilde{\rho}^{n}%
:=\frac{1}{\mathrm{tr}\,A^{n}}A^{n}$. Then,
\begin{align}
\left\Vert \tilde{\rho}^{n}-\rho^{n}\right\Vert _{1}  &  \leq\left\Vert
\frac{1}{\mathrm{tr}\,A^{n}}A^{n}-A^{n}\right\Vert _{1}+\left\Vert A^{n}%
-\rho^{n}\right\Vert _{1}\nonumber\\
&  =\left\vert 1-\mathrm{tr}\,A^{n}\right\vert +\left\Vert A^{n}-\rho
^{n}\right\Vert _{1}\nonumber\\
&  =\left\vert \mathrm{tr}\,\rho^{n}-\mathrm{tr}\,A^{n}\right\vert +\left\Vert
A^{n}-\rho^{n}\right\Vert _{1}\nonumber\\
&  \leq2\left\Vert A^{n}-\rho^{n}\right\Vert _{1}\nonumber\\
&  \leq4\sqrt{2\varepsilon}. \label{datta-1}%
\end{align}
Also,
\begin{equation}
\tilde{\rho}^{n}\leq\frac{2^{na}}{\mathrm{tr}\,A^{n}}\sigma^{n}\leq
\frac{2^{na}}{1-\sqrt{8\varepsilon}}\sigma^{n}. \label{datta-2}%
\end{equation}

\begin{theorem}
\label{th:Dbar=Dmin}%
\[
\overline{\mathrm{D}}\left(  \{\rho^{n}\}||\{\sigma^{n}\}\right)
=\mathrm{D}_{\max}\left(  \{\rho^{n}\}||\{\sigma^{n}\}\right)
\]

\end{theorem}

The proof is analogue of the proof of Theorem\thinspace2 of \cite{Datta}, and
is given below with minor modification in accordance with the difference in
their $\mathrm{D}_{\max}$ and our $\mathrm{D}_{\max}^{\infty}$.

\begin{proof}
First we show `$\leq$'. Let $c>0$ and $\left\{  \tilde{\rho}^{n}\right\}  $ be
a sequence of states with
\begin{align*}
\tilde{\rho}^{n}  &  \leq e^{n\left\{  \mathrm{D}_{\max}\left(  \{\rho
^{n}\}||\{\sigma^{n}\}\right)  +c\right\}  }\sigma^{n},\,\\
\,\lim_{n\rightarrow\infty}\left\Vert \rho^{n}-\tilde{\rho}^{n}\right\Vert
_{1}  &  =0.
\end{align*}
Then, for any $\left\{  P^{n}\right\}  $ with%
\[
\varliminf_{n\rightarrow\infty}\frac{-1}{n}\ln\mathrm{tr}\,P^{n}\sigma^{n}%
\geq\mathrm{D}_{\max}\left(  \{\rho^{n}\}||\{\sigma^{n}\}\right)  +2c,
\]
we have
\begin{align*}
&  \lim_{n\rightarrow\infty}\mathrm{tr}\,P^{n}\rho^{n}=\lim_{n\rightarrow
\infty}\mathrm{tr}\,P^{n}\tilde{\rho}^{n}\\
&  \leq\lim_{n\rightarrow\infty}e^{n\left\{  \mathrm{D}_{\max}\left(
\{\rho^{n}\}||\{\sigma^{n}\}\right)  +c\right\}  }\mathrm{tr}\,P^{n}\sigma
^{n}=0.
\end{align*}
Since $c>0$ is arbitrary, by the second identity of Theorem\thinspace
\ref{th:NH-converse}, we have `$\leq$'.

Second, we show $\geq$`'. Let $a=$ $\overline{\mathrm{D}}\left(  \{\rho
^{n}\}||\{\sigma^{n}\}\right)  +c$ ($c>0$). Then,
\[
\lim_{n\rightarrow\infty}\mathrm{tr}\,\rho^{n}\left\{  \rho^{n}-e^{na}%
\sigma^{n}\leq0\right\}  =1.
\]
Hence, by (\ref{datta-1}) and (\ref{datta-2}), we can compose $\tilde{\rho
}^{n}$ with
\[
\lim_{n\rightarrow\infty}\left\Vert \tilde{\rho}^{n}-\rho^{n}\right\Vert
_{1}=0,\,\,\tilde{\rho}^{n}\leq e^{n\left(  a+c\right)  }\sigma^{n}%
,\,\,\forall c>0\,\exists n_{0}\,\forall n\geq n_{0},
\]
or
\[
\overline{\mathrm{D}}\left(  \{\rho^{n}\}||\{\sigma^{n}\}\right)
+2c\geq\mathrm{D}_{\max}\left(  \{\rho^{n}\}||\{\sigma^{n}\}\right)  .
\]
Since $c$, $c^{\prime}>0$ are arbitrary, we have the assertion.
\end{proof}

\begin{theorem}
\label{th:Dbar=D}\cite{Nagaoka:1999}\cite{NagaokaHayashi}%
\[
\mathrm{D}\left(  \rho||\sigma\right)  =\overline{\mathrm{D}}\left(
\{\rho^{\otimes n}\}||\{\sigma^{\otimes n}\}\right)  .
\]

\end{theorem}

\begin{corollary}
\label{cor:Dmin=D}%
\[
\overline{\mathrm{D}}\left(  \{\rho^{\otimes n}\}||\{\sigma^{\otimes
n}\}\right)  =\mathrm{D}_{\max}\left(  \{\rho^{\otimes n}\}||\{\sigma^{\otimes
n}\}\right)  =\mathrm{D}\left(  \rho||\sigma\right)  .
\]

\end{corollary}

\subsection{Asymptotically lower continuous and monotone relative entropy}

\label{subsec:asym-cont-monotone-relative-entropy}

\textbf{Theorem 2.7} \textit{If }$\mathrm{D}\left(  \rho_{0}||\sigma
_{0}\right)  >\mathrm{D}\left(  \rho||\sigma\right)  $\textit{, there is a
sequence }$\left\{  \Psi^{n}\right\}  $\textit{\ of TPCP map with }%
\begin{equation}
\lim_{n\rightarrow\infty}\left\Vert \Psi^{n}\left(  \rho_{0}^{\otimes
n}\right)  -\rho^{\otimes n}\right\Vert _{1}=0,\,\,\,\Psi^{n}\left(
\sigma_{0}^{\otimes n}\right)  =\sigma^{\otimes n}.\label{asym-def-2}%
\end{equation}
\textit{Conversely, if }$\left\{  \Psi^{n}\right\}  $\textit{\ with
(\ref{asym-def-2}) exists, }$\mathrm{D}\left(  \rho_{0}||\sigma_{0}\right)
\geq\mathrm{D}\left(  \rho||\sigma\right)  $.

\begin{proof}
Suppose $\mathrm{D}\left(  \rho_{0}||\sigma_{0}\right)  >\mathrm{D}\left(
\rho||\sigma\right)  $ and let
\[
c:=\frac{1}{2}\left\{  \mathrm{D}\left(  \rho_{0}||\sigma_{0}\right)
-\mathrm{D}\left(  \rho||\sigma\right)  \right\}  .
\]
By Theorem\thinspace\ref{th:Hiai-Petz}, there is a sequence of projector
$\left\{  P^{n}\right\}  $ with%
\begin{align*}
p^{n}\left(  0\right)   &  :=\mathrm{tr}\,P^{n}\rho_{0}^{\otimes n}%
\rightarrow1\,\left(  n\rightarrow\infty\right) \\
q^{n}\left(  0\right)   &  :=\mathrm{tr}\,P^{n}\sigma_{0}^{\otimes n}\leq
e^{-n\left(  \mathrm{D}\left(  \rho_{0}||\sigma_{0}\right)  -c\right)
}=e^{-n\left(  \mathrm{D}\left(  \rho||\sigma\right)  +c\right)
}\,,\,\,\,\exists n_{0}\forall n\geq n_{0}.
\end{align*}
Let CPTP map $\Phi^{n}$ as of \ref{th:asym-reverse-test}. Then, due to
$\mathrm{D}_{\max}\left(  \{\rho^{\otimes n}\}||\{\sigma^{\otimes n}\}\right)
=\mathrm{D}\left(  \rho||\sigma\right)  $, the composition $\Psi^{n}$ of the
measurement $\left\{  P^{n},\mathbf{1}-P^{n}\right\}  $ followed by $\Phi^{n}$
satisfies (\ref{asym-def-2}). Thus we have the former half of the assertion.

In the sequel, we prove the latter half. Recall $\mathrm{D}\left(
\rho||\sigma\right)  $ satisfies (M), (A), and (C). Therefore,
\begin{align*}
&  \mathrm{D}\left(  \rho_{0}||\sigma_{0}\right)  \underset{\text{(A)}}%
{=}\varliminf_{n\rightarrow\infty}\frac{1}{n}\mathrm{D}\left(  \rho
_{0}^{\otimes n}||\sigma_{0}^{\otimes n}\right) \\
&  \underset{\text{(M)}}{\geq}\varliminf_{n\rightarrow\infty}\frac{1}%
{n}\mathrm{D}\left(  \Psi^{n}\left(  \rho_{0}^{\otimes n}\right)  ||\Psi
^{n}\left(  \sigma_{0}^{\otimes n}\right)  \right) \\
&  =\varliminf_{n\rightarrow\infty}\frac{1}{n}\mathrm{D}\left(  \tilde{\rho
}^{n}||\sigma^{\otimes n}\right)  \underset{\text{(C)}}{\geq}\varlimsup
_{n\rightarrow\infty}\frac{1}{n}\mathrm{D}\left(  \rho^{\otimes n}%
||\sigma^{\otimes n}\right)  \underset{\text{(A)}}{=}\mathrm{D}\left(
\rho||\sigma\right)  .
\end{align*}

\end{proof}

\begin{corollary}
$\mathrm{D}^{F}\left(  \rho||\sigma\right)  :=\ln\left\Vert \sqrt{\rho}%
\sqrt{\sigma}\right\Vert _{1}$ does not satisfy the condition (C).
\end{corollary}

\begin{proof}
$\mathrm{D}^{F}\left(  \rho||\sigma\right)  $ satisfies (M) and (A), but does
not equal a constant multiple of $\mathrm{D}\left(  \rho||\sigma\right)  $.
Therefore, we must have the assertion. \bigskip
\end{proof}

\begin{theorem}
If $g$ is a properly normalized monotone metric, then%
\begin{align*}
&  \inf\left\{  \varliminf_{n\rightarrow\infty}\frac{1}{n}\mathrm{D}%
^{g}\left(  \tilde{\rho}^{n}||\sigma^{\otimes n}\right)  \,;\lim
_{n\rightarrow\infty}\left\Vert \tilde{\rho}^{n}-\rho^{\otimes n}\right\Vert
_{1}=0\right\} \\
&  =\mathrm{D}\left(  \rho||\sigma\right)  .
\end{align*}

\end{theorem}

\begin{proof}
By Theorem\thinspace\ref{th:D<DQ<DR}, we have to prove the assertion only for
$g=J^{R}$. `$\geq$' is due to $\mathrm{D}^{R}\geq\mathrm{D}$ and
Proposition\thinspace\ref{prop:D-asym-cont}. To prove `$\leq$', let $\left(
\Phi^{n},\left\{  p^{n},q^{n}\right\}  \right)  $ be an asymptotic reverse
test with
\[
\varliminf_{n\rightarrow\infty}\frac{-1}{n}\ln q^{n}\left(  0\right)
=\mathrm{D}\left(  \rho||\sigma\right)  +c.
\]
Then by monotonicity of \thinspace$\mathrm{D}^{R}$,
\begin{align*}
&  \varliminf_{n\rightarrow\infty}\frac{1}{n}\mathrm{D}^{R}\left(  \Phi
^{n}\left(  p^{n}\right)  ||\sigma^{\otimes n}\right)  \leq\varliminf
_{n\rightarrow\infty}\frac{1}{n}\mathrm{D}^{R}\left(  p^{n}||q^{n}\right) \\
&  =\varliminf_{n\rightarrow\infty}\frac{1}{n}\mathrm{D}\left(  p^{n}%
||q^{n}\right) \\
&  =\varliminf_{n\rightarrow\infty}\frac{-1}{n}\ln q^{n}\left(  0\right)
=\mathrm{D}\left(  \rho||\sigma\right)  +c,
\end{align*}
which leads to the assertion.
\end{proof}

\section{Conclusions and Discussions{}}

Using reverse test and asymptotic reverse test, we gave a characterization of
quantum versions of relative entropy. Note that the uniqueness in the
asymptotic scenario is valid also for classical relative entropy : any
two-point functions over probability distribution with (A), (M) and (C) is
constant multiple of relative entropy.

The condition (C) can be replaced by the following `weak
monotonicity'\thinspace\cite{Hwang-Matsumoto}, which may be a bit more natural.

\begin{description}
\item[(WM)] (weak monotonicity) If $\left\Vert \tilde{\rho}^{\otimes
n}-\Lambda^{n}\left(  \rho^{\otimes n}\right)  \right\Vert \rightarrow0$,
$\tilde{\sigma}^{\otimes n}=\Lambda^{n}\left(  \sigma^{\otimes n}\right)  $
\[
\mathrm{D}^{Q}\left(  \rho||\sigma\right)  \geq\mathrm{D}^{Q}\left(
\tilde{\rho}||\tilde{\sigma}\right)  .
\]

\end{description}

It may be interesting to compare the asymptotic behavior of quantum relative
entropy and corresponding quantum Fisher information (correspondence is made
via Lemma\thinspace\ref{lem:D-int} ). While it is known that $J^{R}$ and
$J^{S}$ satisfies both of them\thinspace\cite{Matsumoto:2010}, $\mathrm{D}%
^{R}\left(  \rho||\sigma\right)  $ and $\mathrm{D}^{J^{S}}\left(  \rho
||\sigma\right)  $ does not satisfy (C) and (A), respectively.

Some problems are left open. First, relaxing (C) in the following manner can
be interesting :

\begin{description}
\item[(C')] (Lower exponential asymptotic continuity) If $\ \ \varliminf
_{n\rightarrow\infty}\frac{-1}{n}\ln\left\Vert \tilde{\rho}^{n}-\rho^{\otimes
n}\right\Vert \geq a$, \
\[
\varliminf_{n\rightarrow\infty}\frac{1}{n}\left\{  \mathrm{D}^{Q}\left(
\tilde{\rho}^{n}||\sigma^{\otimes n}\right)  -\mathrm{D}^{Q}\left(
\rho^{\otimes n}||\sigma^{\otimes n}\right)  \right\}  \geq0.
\]

\end{description}

By relaxing (C) to (C'), quantities such as relative Renyi entropy may
survive. Second, generalizing Theorem\thinspace\ref{th:asym-reverse-test} and
Theorem\thinspace\ref{th:asym-q-template} (by increasing the numbers of
states, changing constraint on error, etc.) may be also interesting. \

\section{Reverse estimation for a multi-dimensional parameter family}

\label{appendix:multi-reverse-est}

\bigskip By the argument parallel with the $1-\dim$- case, we have
\[
J_{\rho_{\theta}}^{R}\left(  X,X\right)  =\min\left\{  J_{p_{\theta}}\left(
Y,Y\right)  ;\Psi\left(  p_{\theta}\right)  =\rho_{\theta},\Psi\left(
Y\right)  =X\right\}  .
\]
Therefore, for any reverse estimation $\left(  \Phi,\left\{  p_{\theta
}\right\}  \right)  $,
\[
J_{p_{\theta}}\leq J_{\rho_{\theta}}^{R},
\]
which, for any real $m\times m$-matrix $G>0$, leads to%

\begin{align}
\mathrm{Tr}\,GJ_{\theta}  &  \geq\min\left\{  \mathrm{Tr}GJ\,;\,J\geq
J_{\theta}^{R}\right\}  =\mathrm{Tr}\,G\Re J_{\theta}^{R}+\mathrm{Trabs}\,G\Im
J_{\theta}^{R}\nonumber\\
&  =\mathrm{Tr}\,G\Re J_{\theta}^{R}+\mathrm{Trabs}\,G\Im J_{\theta}^{R}
\label{trGJ>trGJR}%
\end{align}
Note that
\begin{align*}
\Im J_{\theta,ij}^{R}  &  =\frac{1}{2}\left(  \mathrm{Tr}\rho_{\theta
}L_{\theta,i}^{R\dagger}L_{\theta,j}^{R}-\mathrm{Tr}L_{\theta,j}^{R\dagger
}L_{\theta,i}^{R}\rho_{\theta}\right)  =\frac{1}{2}\left(  \mathrm{Tr}%
L_{\theta,i}^{R}\rho_{\theta}L_{\theta,j}^{R}-\mathrm{Tr}L_{\theta,i}%
^{R}L_{\theta,j}^{R}\rho_{\theta}\right) \\
&  =-\frac{1}{2}\mathrm{Tr}\rho_{\theta}\left[  L_{\theta,i}^{R}%
,\,L_{\theta,j}^{R}\right]  :=\tilde{J}_{\theta,ij},
\end{align*}
This inequality is in many cases not achievable. \ However, if $\left\{
\rho_{\theta}\right\}  $ is RLD-parallel, $\Im J_{\theta,ij}^{R}=0$ and the
inequality is written as
\[
\mathrm{Tr}\,GJ_{\theta}\geq\mathrm{Tr}\,G\Re J_{\theta}^{R},
\]
which is achievable. Also:

\begin{example}
Gaussian states are defined by its P-representation,
\[
\rho_{\theta}=\int\frac{\mathrm{d}p\mathrm{d}q}{2\pi\overline{N}}\exp\left[
-\frac{(q-\theta^{1})^{2}+(p-\theta^{2})^{2}}{2\overline{N}}\right]
\left\vert \frac{q+ip}{\sqrt{2}}\right\rangle \left\langle \frac{q+ip}%
{\sqrt{2}}\right\vert
\]
where $\left\vert z\right\rangle $ is the coherent state with complex
amplitude $z$. Being infinite dimensional states, in strict sense, this
example is out of the scope of our theory. However, the lower bound
(\ref{trGJ>trGJR}) can be explicitly computed as
\[
\mathrm{Tr}\Re J_{\theta}^{R}+\mathrm{Trabs}\Im J_{\theta}^{R}=\frac
{2}{\overline{N}}.
\]
Also, using P-representation, one can compose a reverse estimation (generate
coherent states according to the probability distribution defined by
$P$-function), and
\[
J=\left(
\begin{array}
[c]{cc}%
\overline{N} & 0\\
0 & \overline{N}%
\end{array}
\right)  ,\,\mathrm{Tr}\,J=\frac{2}{\overline{N}},
\]
achieving the lower bound.

Note that the optimal measurement for estimation of $\theta=\left(  \theta
^{1},\theta^{2}\right)  $ is $\left\{  \left\vert \frac{\hat{\theta}^{1}%
+i\hat{\theta}^{2}}{\sqrt{2}}\right\rangle \left\langle \frac{\hat{\theta}%
^{1}+i\hat{\theta}^{2}}{\sqrt{2}}\right\vert \right\}  $ \cite{Holevo}, and
the distribution of the estimate of $\theta=\left(  \theta^{1},\theta
^{2}\right)  $ is equal to Q-representation. Moreover, let
\[
\left\vert \varphi_{\theta}\right\rangle :=\sum_{n=0}^{\infty}\left(
\frac{\overline{N}}{\overline{N}+1}\right)  ^{\frac{n}{2}}U_{\theta}\left\vert
n\right\rangle \otimes U_{\theta}\left\vert n\right\rangle \in\mathcal{H}%
\otimes\mathcal{K},
\]
where $U_{\theta}$ is the Weyl operator. Then $\rho_{\theta}=\mathrm{tr}%
_{\mathcal{K}}\,\left\vert \varphi_{\theta}\right\rangle \left\langle
\varphi_{\theta}\right\vert =\mathrm{tr}_{\mathcal{H}}\,\left\vert
\varphi_{\theta}\right\rangle \left\langle \varphi_{\theta}\right\vert $ and
\[
\left\langle 2^{-\frac{1}{2}}\left(  \hat{\theta}^{1}+i\hat{\theta}%
^{2}\right)  \right.  \left\vert \varphi_{\theta}\right\rangle =\sqrt{\frac
{1}{\overline{N}+1}}\exp\left\{  -\frac{(\hat{\theta}^{1}-\theta^{1}%
)^{2}+(\hat{\theta}^{2}-\theta^{2})^{2}}{4\left(  \overline{N}+1\right)
}\right\}  \left\vert \sqrt{\frac{\overline{N}}{\overline{N}+1}}\frac
{\hat{\theta}^{1}-i\hat{\theta}^{2}}{\sqrt{2}}\right\rangle .
\]
Therefore, if one measures $\mathcal{K}$-part of $\left\vert \varphi_{\theta
}\right\rangle $ by POVM $\left\{  \left\vert \frac{\hat{\theta}^{1}%
+i\hat{\theta}^{2}}{\sqrt{2}}\right\rangle \left\langle \frac{\hat{\theta}%
^{1}+i\hat{\theta}^{2}}{\sqrt{2}}\right\vert \right\}  $, one obtains
$\left\vert \frac{\hat{\theta}^{1}-i\hat{\theta}^{2}}{\sqrt{2}}\right\rangle $
with probability $\frac{1}{2\pi\overline{N}}\exp\left\{  -\frac{(\hat{\theta
}^{1}-\theta^{1})^{2}+(\hat{\theta}^{2}-\theta^{2})^{2}}{2\overline{N}%
}\right\}  $, or realizes the optimal reverse estimation.
\end{example}

\end{document}